# Increase of Thermal Resistance Between a Nanostructure and a Surface due to Phonon Multireflections

**Sebastian Volz[1] and Pierre-Olivier Chapuis**

Laboratoire d'Energétique Moléculaire et Macroscopique, Combustion, UPR CNRS 288

Ecole Centrale Paris, Grande Voie des Vignes

92295 Châtenay Malabry, France

**ABSTRACT:** The thermal resistance between a nanostructure and a half-body is calculated in the framework of particle-phonons physics. The current models approximate the nanostructure as a thermal bath. We prove that the multireflections of heat carriers in the nanostructure significantly increase resistance in contradiction with former predictions. This increase depends on the shape of the nanostructure and the heat carriers mean free path only. We provide a general and simple expression for the contact resistance and examine the specific cases of nanowires and nanoparticles.

---

[1] Corresponding author: volz@em2c.ecp.fr

# 1. Introduction

Fourier heat conduction model is not relevant on the nanoscale because the involved dimensions are smaller or comparable to the heat carriers mean free path. Drastic deviations are encountered,[1-3] and new approaches have to be elaborated. Fourier law is unable to predict the heat flux in cases where the size of the structure, the heat source or the thermal heterogeneities on the boundaries are on the order of magnitude or smaller than the phonon mean free path. Under such conditions the heat transport is partially ballistic: heat carriers rarely interact in the volume of interest.

We address the thermal resistance between a nanostructure and a half-body when the contact also has a small characteristic length. The current model[4] describing nanocontacts introduces a correction to the case of a macrocontact. However, it still assumes Fourier heat conduction not only in the half-body but also in the nanostructure. This is because the contact size is considered smaller than the characteristic size of the nanostructure. In this situation, phonons coming in the nanostructure have a very low probability of coming back to the contact. They thermalize in the nanostructure that is therefore assumed as fully absorbing as a heat bath. This situation is shown in Fig. 1a.

However, since a nanostructure is commonly defined by a characteristic size between 10nm and 500 nm, the contact cross section has to be much smaller than 10 - 500nm to ensure the condition of fully absorbing heat bath. Under these constraints, phonon particle physics is not relevant anymore because the wavelengths of the thermal phonons are of the same order of magnitude as the contact size, making wave effects, such as phonon diffraction, become significant.

For a nanosized structure, we believe that the particle-phonon approximation can only provide information when the contact dimension is on the same order of magnitude as the structure dimension. Besides, the nanostructure has also a characteristic dimension on the same order of magnitude than the typical mean free path in crystals. As illustrated in Fig. 1b, the consequence is that phonons are reflected on the nanostructure surface and eventually return to the half body: the current model does not apply anymore.

The objective of our study is to understand and predict the impact of phonon reflections on the thermal resistance.* Our predictions reveal that this thermal resistance can be enhanced several times compared to the ones of current descriptions.

This objective is relevant to several applications such as (i) nanocontacts between a low dimensional structure (nanowire, nanotube, nanoparticle) and a surface,[4] (ii) fabrication processes such as nanolithography,[5] (iii) any nano/microscale thermal measurements based on contact probes[6,7] and (iv) interfacial thermal resistance where the solid-solid micro/nanocontacts cause constrictions of the heat flux lines in both materials.[8,9]

Section 2 presents the physical model that starts from the current theory and proposes a general treatment of the nanostructure/surface resistance. The framework is based on the assumption that the transport regime in the half-body is Fourier like. Results of calculations are reported and explained in the first part of Section 3. In the second part of Section 3, we estimate the deviation due to the non-Fourier regime in the half-body.

**2. Physical Model**

**2.1. Nanocontact between two thermal baths**

---

* Note that our approach is based on the analogy between phonons and photons. A clear introduction on radiation principles is provided by reference 16.

Our physical model is based on the work by Nikolic and Allen.[10] They proposed an analytical calculation of the electrical resistance between two reservoirs. The two bodies are linked by a circular constriction. We consider their model in the framework of heat transfer where electron reservoirs are replaced by thermal baths of phonons. Wexler[11] proposed an approximated calculation for approaching the exact solution. This approximation is formulated as the sum of diffusive and ballistic resistance.

Maxwell theory[12] applied to heat transfer yields the diffusive resistance $R_M=1/(Dk)$ where k is the thermal conductivity and D is the contact size. This resistance is the sum of the resistances created by two thermal baths. They are due to the constrictions of the heat flux lines in the vicinity of the contact.

The ballistic resistance is proportional to the reverse of the phonon heat flux through the contact cross section. However, predominant phonon scattering is due to the interaction between phonons and the perimeter of the contact instead of the interaction between phonons. The relevant scattering length is now proportional to the contact size D. The contact acts as a bottleneck. This ballistic resistance is known as the Sharvin[13] term in electronics and does not depend on phonon-phonon scattering or mean free path.

The resulting thermal resistance between two thermal baths linked by a circular contact can finally be written as:

$$R_W = \frac{1}{kD} + \frac{16}{\pi CvD^2} = \frac{3}{CvD}\left(\frac{1}{\Lambda} + \frac{16}{3\pi D}\right), \qquad (1)$$

In this equation, $\Lambda$ is the mean free path, C and v are the phonons volumetric capacity and average group velocity respectively. We have considered the Debye expression of the thermal conductivity $k=Cv\Lambda/3$ to derive the RHS term. The RHS term includes a phonon mean free path as defined by the following Matthiessen rule $(1/\Lambda+16/3\pi D)^{-1}$. When the contact size D is much larger than the phonon-phonon mean free path Λ, the resulting mean free path equals to Λ and the Maxwell resistance is retrieved

When factoring the mean free path Λ in the denominator of Eq. (1), the dimensionless Knudsen number Kn=Λ/D appears as the key quantity to estimate the deviation to the Maxwell resistance. It was proven[10] that the large and small Knudsen limits predicted by Eq. (1) accurately match the analytical results. However this expression presents a maximal deviation of 11% for Kn=1 when compared to the exact solution.

Eq. (1) implies deep consequences because the resistance $R_W$ becomes independent to the mean free path when the Sharvin term is predominant. This happens as early as when the Knudsen number Kn=Λ/D is larger than 3π/16=0.589. For instance, measuring the thermal conductivity of a sample with a contact probe on a characteristic length smaller than the phonon-phonon mean free path is not feasible. The reason is that the thermal resistance of the sample becomes independent to the mean free path.

**2.2. Nanocontact between a nanostructure and a half body**

We aim at correcting Eq. (1) because it fails to describe the case of a nanostructure/half-body contact. Fourier conduction does not capture the relevant physical mechanisms in the nanostructure.

Figure 2 provides a schematic of the different regimes that occur when the characteristic sizes of the structure L and the contact D are varied. When L and D are large, the classical Maxwell resistance $R_M$ is relevant. When the structure dimension L is larger than the mean free path but D is smaller than the mean free path, the structure is assimilated as a perfect phonon absorber and the Wexler formula (dotted background) applies.

However, the Wexler formula is not adequate to describe the case of nanostructures because the hypothesis of a perfect phonon absorber implies that L>>D. On the other hand, a nanostructure is typically smaller than 500nm, and the contact size, in turn, has to be smaller than 10nm. A wavelike behaviour of phonons is expected at such small dimensions but it is not included in Wexler formula.

Our work focuses on the case where the characteristic dimensions L, D and Λ are on the same order of magnitude. But the schematic of Fig. 2 also shows that our work does not address the wavelike behaviour of phonons such as phonon transmission (background with hatchings). Recent works have investigated this effect in the case of constrictions between nanospheres.[14] We next explain how to model the impact of the nanostructure by correcting Eq.(1).

**2.3. Defining thermal resistance**

The flux and the temperature difference between the half-body and the nanostructure are sufficient to define the thermal resistance. The thermal bath allows for definition of the temperature $T_0$ away from the aperture. But the second reference temperature is more difficult to identify. The temperatures in the nanostructure and in the vicinity of the contact are ill-defined quantities because non-equilibrium heat transfer is involved. To define a second temperature reference, we assume that the nanostructure is coupled to an external thermal bath at temperature $T_1$. In practical conditions, the coupling can be radiative, it can be done by forced convection or even by conduction through air, water or solid contact.

**2.4. The contribution of the half body**

The heat transfer in the half body is Fourier like at remote distances from the contact. The half of the Maxwell resistance $R_F=R_M/2=1/(2Dk)$ then accounts for the constrictions of the heat flux lines in this region. Nearer to the contact, a partially ballistic heat transfer is expected. We already noted that the deviation from the Wexler formula in Eq. (1) is due to this partially ballistic regime and remains smaller than 11%. For a first approximation, we will neglect the influence of this regime and propose a general and simple expression that accounts for the nanostructure. Later, in Section 3.2, we will provide a correction to the previous approximation.

**2.5. Defining non-equilibrium effective temperatures in the nanostructure**

In the nano-object, an equilibrium phonon distribution at $T_1$ is superimposed to the incoming phonons at temperature $T_0$. Those heat carriers interact with themselves and with the nanostructure surface but they undergo a low number of scattering events: they cannot thermalize. The resulting phonon distribution is hence characterized by a non-equilibrium or a non-Fourier regime. This regime can be treated by calculating heat fluxes but we introduce effective temperatures to interpret the deviation to the thermodynamic temperature $T_1$ used in Wexler formula of Eq.(1).

Firstly, we define the temperature of emission T that is related to the non-Fourier heat flux q according to the following expression:

$$q = \frac{1}{2\pi} \int_{\omega,\Omega_{2\pi}} g_{2\pi}(\omega)\hat{E}(\Omega)\hat{E}\cos\theta \ \hbar\omega \ f(\Omega) \, d\omega \, d\Omega$$

$$= \frac{\hbar}{8\pi^2 v^2} \int_{\omega=0}^{\omega_D} \frac{\omega^3}{\exp\left(\frac{\hbar\omega}{k_B T}\right)-1} \, d\omega = \frac{\omega_D^3 \ k_B}{24\pi^2 v^2} T \qquad (2)$$

$k_B$ is the Boltzmann constant, $\omega$ represents the phonon angular frequency and $\omega_D$ is the Debye angular frequency. g corresponds to the phonon density of states which is expressed according to the Debye approximation and the group velocity is isotropic and frequency independent as postulated by the same approximation. The index 2π refers to the directions of 2π steradians. cosθ indicates that the velocity is projected on the direction perpendicular to the surface, θ being the angle between the phonon velocity and the direction perpendicular to the surface.

Eq. (2) is the general expression of a phonon heat flux but here, f is the number of phonons coming from the direction Ω and f is not isotropic. This reveals the non-equilibrium transport. As shown in Eq. (2), we assume that f can be related to an isotropic Bose-Einstein distribution including an effective temperature T. This approximation is not that crude because the variation of the quantity f along directions remains small and the Bose Einstein distribution is an average over directions of those variations. The temperature T is set larger than the Debye temperature so that the flux is finally proportional to T.

In the contact cross section, Eq. (2) defines the emission temperature $T_D^i$ related to the heat flux that is emitted from the nanostructure towards the half-body. The superscript i refers to an incident flux and the index D corresponds to the contact surface $S_D$.

Another type of effective temperature can also be calculated from the local energy density as follows:

$$\frac{1}{4\pi} \int_{\omega,\Omega_{4\pi}} g_{4\pi}(\omega)\hbar\omega \, f(\Omega) \, d\omega \, d\Omega = \frac{\hbar}{4\pi^2 v^3} \int_{\omega=0}^{\omega_D} \frac{\omega^3}{\exp\left(\frac{\hbar\omega}{k_B T}\right) - 1} \, d\omega = \frac{\omega_D^3 \, k_B}{12\pi^2 v^3} T. \quad (3)$$

Once again, the temperature T that can be compared to an effective thermodynamic temperature, defines an average over directions of the phonon number f. The index 4π refers to the directions of 4π steradians. Eq. (3) allows for deriving the expression of the effective thermodynamic temperature $T_a$ in the contact cross section. $T_a$ is estimated as the algebraic average of the temperatures $T_D^i$ and $T_0$ because the f function is a Bose-Einstein distribution at temperature $T_D^i$ in the directions of one hemisphere and at temperature $T_0$ in the directions of the other hemisphere.

$T_D^i$ and $T_a$ refer to a heat flux and an energy respectively. Using those temperatures will allow us to calculate the deviation to the temperature $T_1$ due to the non-Fourier regime.

**2.6. Expression of the contact resistance**

Our strategy is to correct the Sharvin resistance of the contact and the resistance $R_F$ associated with one thermal bath. When the nanostructure replaces the second thermal bath, the temperature difference defining the net heat flux through the contact is not $(T_1-T_0)$ but $(T_D^i-T_0)$. Between the contact and the thermal bath, the relevant temperature difference is not $(T_1-T_0)/2$ anymore but $(T_a-T_0)$. We will show that the resistance R defining the heat flux with $(T_1-T_0)$ as reference, is obtained by the following relation:

$$q = \frac{R}{(T_1 - T_0)} = \frac{R_W}{(T_D^i - T_0)}, \qquad (4)$$

where $R_W$ is the Wexler resistance defined in Eq. (1). The correction coefficients to the resistance $R_F$ and the Sharvin resistances appear to be the same, this correcting coefficient is the temperature ratio $(T_1-T_0)/(T_D^i-T_0)$.

We now provide an analytical expression of this ratio. After a thorough derivation including the coupling with a thermal bath at temperature $T_1$ as well as the phonon-phonon and surface scattering in the nanostructure (APPENDIX I), we express the ratio as follows:

$$\frac{T_D^i - T_0}{T_1 - T_0} = 1 - \gamma, \qquad (5)$$

The coefficient $\gamma = \dfrac{\tau_{1D}^2}{1-\tau_{11}}$ introduces the geometric-mean transmittances

$\tau_{1i} = (\pi S_1)^{-1} \displaystyle\int_{\Omega(S_1), S_i} e^{-\frac{r}{\Lambda}} \mathbf{u}.d\mathbf{S}_i d\Omega$ where the indexes 1 and D (index i=1 or D), refer to the surfaces $S_1$ and $S_D$ of the nanostructure and of the contact respectively. The scattering is treated along paths having lengths described by the variable r and the direction **u**. Those paths link the surface element $dS_i$ to the surface element $dS_1$. $d\Omega$ is the element of solid angle.

The geometric-mean transmittance $\tau_{1i}$ is the fraction of the heat flux leaving surface 1 and reaching surface i after several phonon-phonon scattering. $\gamma$ is the fraction of the heat flux leaving the nanostructure and carrying phonons at temperature $T_0$. This term is proportional to the heat flux leaving the surface $S_D$ towards the thermal bath. This flux is proportional to $\dfrac{\tau_{1D}}{1-\tau_{11}}$ as shown in APPENDIX I and is attenuated by phonon-phonon scattering before reaching the surface $S_D$. This scattering is modelled by multiplying the ratio $\dfrac{\tau_{1D}}{1-\tau_{11}}$ by $\tau_{1D}$.

In the ballistic regime, i.e. when $L \ll \Lambda$ and $e^{-r/\Lambda}=1$, the transmittance is equal to its upper limit that is called configuration factor $\alpha_{1i}$. The quantity $\alpha_{1i}$ is defined when no scattering occurs. It is equal to the flux leaving the surface $S_1$ and reaching the surface $S_i$ divided by the total heat flux leaving the surface $S_1$. The heat flux balance yields to $\alpha_{11}+\alpha_{1D}=1$ and finally $\gamma=\alpha_{1D}$ when neglecting scattering. Note that the geometric-mean transmittance $\tau_{1i}$ and $\gamma$ can be computed for any structure shape from commercial heat transfer codes including semi-transparent radiation.

The correction to the Sharvin term consists of replacing $T_1$ by the effective temperature $T_D^i$ but the resistance $R_F$ is also affected by the nanostructure. The correction for the temperature difference defining the heat flux in the half body is derived as follows:

$$\frac{T_a - T_0}{(T_1 + T_0)/2 - T_0} = 1 - \gamma, \quad (6)$$

which is the same as for the Sharvin term. Eq. (6) arises from the calculation of $T_a$ as the algebraic average of $T_D^i$ and $T_0$ (APPENDIX I):

$$\frac{T_a - T_0}{T_1 - T_0} = \frac{1 - \gamma}{2}. \quad (7)$$

Finally, Eq. (1) can be generalized by dividing both the resistance of one thermal bath and the Sharvin resistance by 1-$\gamma$. This is Eq. (4) and it can be expressed by normalizing the contact resistance R by the resistance $R_F$ to yield:

$$\frac{R}{R_F} = \frac{1}{1 - \gamma}(1 + \beta\ Kn) \quad (8)$$

The factor $\beta = 4D/(3R_F S_D k)$ is a non-dimensioned figure accounting for the shape of the contact: $\beta=3.395$ for the disk of diameter D, $\beta=0.59$ for the square of edge D, and $\beta=2.24$ for the line of width D. $\beta$ is easily derived from a classical heat conduction model. Note that Eq. (8) holds for any shape of nanostructure and contact.

When considering different materials in the half-body and in the nano-object, the second RHS term of Eq. (8), i.e. $\beta Kn$, has to be divided by the phonon transmission coefficient from the half-body to the nanostructure. The $\gamma$ coefficient also has to include the phonon mean free path of the nanostructure whereas $R_F$ depends on the phonon mean free path in the half body.

A direct consequence of Eq. (8) is that the thermal resistance is significantly enhanced when $\gamma$ goes to 1. Under these circumstances, $\tau_{1D}^2 + \tau_{11}$ also becomes one, which corresponds to the case of a ballistic regime in the nanostructure. The contact resistance also becomes very large because the nanostructure reflects all the phonons at temperature $T_0$ back to the half body without absorbing their energy.

## 3. Results and discussion

### 3.1. Multireflections in the nanostructure

255

To prove the significant weight of multireflections in the nanostructure, we have calculated the ratio $R/R_F$ of Eq. (8) in four cases: the strip, the wire perpendicular to the surface, the wire lying on the surface and the dot. Although a precise numerical calculation of γ is possible without technical difficulty, we propose a direct estimation of γ based on the geometric-mean

260  beam length approximation. In this framework, the geometric-mean transmittance $\tau_{1i}$ is assumed to be equal to $\alpha_i(1-L_{1i}/\Lambda)$ where $L_{1i} = \dfrac{1}{S_1 \alpha_{1i}} \int_{S_1} \int_{S_i} \dfrac{d\mathbf{S}_1.\mathbf{n}_1 \; d\mathbf{S}_i.\mathbf{n}_i}{\pi.r}$ is the geometric-mean beam length.[15] $\mathbf{n}_1$ and $\mathbf{n}_i$ are the unit vectors with directions parallel to the vector $\mathbf{r}$ that is joining both surface elements. The previous expression of $\tau_{1i}$ imposes $L_{1i}<\Lambda$ which is confirmed in three of the four cases when Kn>1.

265

The detailed derivations of the γ coefficients are provided in APPENDIX II and they are reported in Table 1.

We noted that the Knudsen number must be larger than one for the mean beam length approximation to be applied. Therefore $\tau_{1i}$ is well defined and remains larger than zero except

270  for the wire perpendicular to the surface but the configuration factor $\alpha_{1D}$ goes to zero when the wire length increases and $\tau_{1i}$ also reduces to zero in this case.

We sought to better understand the impact of the Knudsen number Kn=Λ/D. Therefore, we replaced γ by its expression as a function of Kn and report the resistance deviation $\delta R/R_W$ against the Knudsen number in Fig. 3a. δR represents the difference between the corrected

275  resistance R of Eq. (8) and the one predicted by the Wexler approximation in Eq. (1).

In the case of the strip geometry, Figure 3a reveals an enhancement of the thermal resistance by a factor of five when Kn=5. This difference remains significant even when Kn=1 because the contact resistance is still twice larger than in the half-body/half-body case. We envisioned a strong impact of this result on the heat transfer of integrated circuits (ICs). The phonon mean free path in silicon is equal to 100nm and the metal tracks of ICs have widths in the same range. The geometry of a track is comparable to that of the strip presented above. The increased thermal resistance between the track and the substrate might generate a significant temperature rise in and just below the track.

For Kn=5, the data obtained with the other geometries also indicate a resistance enhancement of 12% (cube) and 27% (wire deposited on the surface). The deviation for the nanowire grown perpendicular to the surface remains negligible as it behaves like a phonon absorber. When the Knudsen number increases to higher values, the resistance deviation for the strip reaches arbitrarily large values. The deviation reaches an asymptotic value of 50% for the horizontal nanowire and of 25% for the cube. These figures are predicted by the ballistic limit of the ratio $\delta R/R_W = \gamma/(1-\gamma) = \alpha_{1D}/(1-\alpha_{1D})$. In this limit, $\delta R/R_W$ only depends on the surface ratio $S_1/S_D$ according to the expression $\delta R/R_W = 1/\left[(S_1/S_D)-1\right]$ because a trivial derivation yields $\alpha_{1D}=S_D/S_1$.[15] The physical meaning of this regime is that the larger the surface $S_1$, the smaller the probability for a phonon to leave the nanostructure. The nano-object then becomes a perfect phonon absorber and the deviation $\delta R$ decreases to zero.

Calculating the resistance ratio $\delta R/R$ leads to the coefficient $\gamma$. This point precisely reveals the physical meaning of $\gamma$, which clearly appears here as the relative deviation of the resistance compared to the Wexler prediction. In Fig. 3b, $\gamma=\delta R/R$ is reported against the Knudsen number. The increase of this last resistance ratio is smaller than the one of $\delta R/R_W$ because R increases more rapidly with Knudsen number than $R_W$.

### 3.2. Partially ballistic regime in the thermal bath

Previous work[10] predicted an 11% deviation of the resistance derived from the Wexler expression of Eq. (7) when compared to the exact thermal resistance. The reason is that Wexler formula is an approximated Matthiessen rule describing the partially ballistic heat transfer in the half body. Finding the general and exact solution of the nanoparticle/half body thermal resistance is an unfeasible task, at least if a rather simple expression is targeted. Here we aim to prove that this deviation between Matthiessen solution and the exact one remains constant whatever the $\gamma$ value is. We will show that the ratio between the exact resistance and the corrected resistance of Eq. (8) does not depend on the $\gamma$ coefficient. Our strategy consists in deriving a linear dependence between the heat flux in the contact cross section and the temperature difference $T_D^i - T_0$.

The proof is based on the Ballistic Diffusive model[15] that allows for solving the Boltzmann transport equation (BTE). This model is analogous to the Modified Differential Approximation for the radiative transfer equation.[16] Derivation of this model starts with the Boltzmann equation under the relaxation time approximation:

$$\frac{\partial f}{\partial t} + \mathbf{v} \cdot \nabla_r f = -\frac{f - f_0}{\tau}, \quad (9)$$

where $f_0$ is the equilibrium number of phonons and $\tau$ the average phonon relaxation time. The ballistic-diffusive approximation consists in dividing the distribution function into two parts $f(\mathbf{r},\mathbf{u}) = f_m(\mathbf{r},\mathbf{u}) + f_b(\mathbf{r},\mathbf{u})$. $f_b(\mathbf{r},\mathbf{u})$ represents the fraction of heat carriers which have been emitted from the boundaries along the direction defined by $\mathbf{u}$ and arriving at $\mathbf{r}$. $f_m(\mathbf{r},\mathbf{u})$ represents the heat carriers density in the vicinity of position $\mathbf{r}$ arriving from the same direction $\mathbf{u}$. The local heat flux $\mathbf{q}$ is the sum of the ballistic and medium fluxes $\mathbf{q}_b$ and $\mathbf{q}_m$ respectively. $f_b(\mathbf{r},\mathbf{u})$ is a solution of the Boltzmann equation when $f_0(\mathbf{r},\mathbf{u}) = 0$:

325 $$f_b(\mathbf{r},\mathbf{u}) = f_w(\mathbf{r}-\mathbf{r}_0) \cdot \exp(-\frac{r}{\Lambda}). \qquad (10)$$

$f_w$ is the carriers density emitted from the boundary point $\mathbf{r}_0$ along the direction $\mathbf{u}$. The BTE written for $f_m$ combined with the energy balance equation yields: [15]

$$\nabla(\mathbf{q}_b - k\nabla T_m) = 0 \qquad (11)$$

The ballistic heat flux can be computed separately by combining Eqs. (2) and (10). The
330 divergence of the ballistic fluxes $\nabla \mathbf{q}_b$ can be derived from Eq. (10) and inserted as a source term in Eq. (11). From this point of view, Eq. (11) remains a classical heat conduction equation with volumetric sources prescribed by $\nabla \mathbf{q}_b$ and with a temperature $T_a$ as boundary condition over the contact cross section. Calculating the heat flux $\mathbf{q}_b$ from Eq. (2) requires setting the temperature $T_D^i$ as boundary condition on the contact cross section. Note that the
335 coupling between the ballistic-diffusive calculation in the half-body and the nanostructure is achieved by applying the above-mentioned boundary conditions.

We emphasize that the ballistic-diffusive equations provide the correct solutions at the ballistic and diffusive limits of high and low Knudsen values.[17] This statement was confirmed by numerical studies in the 1D case.[14] The 1D analysis also reveals a maximum inaccuracy of
340 1.4% when Kn=1.

We now show that the ballistic heat flux $q_b$ is proportional to the temperature difference $T_D^i - T_0$. To demonstrate this dependence, we decompose the expression (2) of the ballistic heat flux into contributions corresponding to different solid angles as follows:

$$q_b(\mathbf{r}) = \frac{1}{4\pi}\int_\omega g_{FS}(\omega)v\hbar\omega \, d\omega \cdot \left[f_b(\mathbf{r},T_0)I(\Omega_{4\pi}) - f_b(\mathbf{r},T_0)I(\Omega_D) + f_b(\mathbf{r},T_D^i)I(\Omega_D)\right]. \qquad (12)$$

345 We have introduced the quantity $I(\Omega) = \int_{\varphi,\theta\in\Omega} e^{-\frac{r'}{\Lambda}} \cos\theta \sin\theta \, d\theta \, d\varphi$ where r' is the distance between the point with coordinates defined by the position vector $\mathbf{r}$ and the boundary point defined by the direction $\Omega$ and the previous position. $\varphi$ denotes the azimuth angle.

The local thermal equilibrium leads to the equality $f_w(T_0)I(\Omega_{4\pi})=0$ because the sum of the heat fluxes coming from all directions in an isothermal cavity should cancel. Following Eq. (2), the two remaining RHS terms in Eq. (12) can be expressed as linearly dependent to the temperatures $T_0$ and $T_D^i$ respectively. The ballistic heat flux $q_b$ finally arises as the product between a geometric term and a term including the energy as follows: $q_b(r) \propto (T_D^i - T_0)I(\Omega_D)$. The proportionality between the heat flux $q_b(r)$ and the temperature difference $T_D^i - T_0$ is hence verified.

In addition, the local thermal equilibrium implies that $\text{div}\mathbf{q}_m(\mathbf{r}) = -\text{div}\mathbf{q}_b(\mathbf{r})$. The divergence operator only acts on the $I(\Omega_D)$ function in such a way that $q_m$ is also proportional to $T_D^i - T_0$. As a consequence, the resulting heat flux $\mathbf{q} = \mathbf{q}_b + \mathbf{q}_m$ is proportional to the temperature difference $T_D^i - T_0$.

When introducing this last temperature difference in the expression of the exact thermal resistance R', it turns out that:

$$R' = \frac{(T_1 - T_0)}{\int_{S_D} (\mathbf{q}_b(r) + \mathbf{q}_m(r)) d\mathbf{S}_D} \propto \frac{(T_1 - T_0)}{(T_D^i - T_0)} \propto \frac{1}{(1-\gamma)} \ . \tag{13}$$

To numerically show this dependence, we have solved Eq. (11) when the contact cross section is a disk of diameter D. The disk heats a half body which is modelled by a cylinder with boundaries at temperature $T_0=300K$. System symmetry around the cylinder axis is assumed. We set the cylinder height and radius to $L_x = 6$ μm and $L_y = 3$ μm. The temperature field is calculated based on a finite volumes method currently used to solve conventional Fourier conduction problems. We choose to set up a regular 100 x 100 grid of ring elements with square sections. To preserve the approximation of semi-infinite body, the Knudsen number $Kn = \Lambda/D$ is defined between 0.1 and 2.5. The value of D is tuned to provide a rather continuous set of resistances versus Kn. The ratio between the mesh size and the diameter D varies and numerical uncertainties also do. We therefore acknowledge a numerical accuracy of 5-10% by computing the same Knudsen value with different set of parameters. The thermal resistance $R_F$ in the diffusive limit is obtained from the heat flux $q_m$ computed when the ballistic heat flux is removed in Eq. (11).

The ratio $R'/R_F$ versus Kn is reported in Figure 4 for two values of γ. The main point is that the quantity $\delta R'/R'$ x (1-γ) is clearly not γ dependent. This result provides a numerical proof of Eq. (13). Computing other cases with different values of γ would basically confirm the dependence of the quantity $R'/R_F$ to the coefficient 1/(1-γ).

To sum up, the exact solution for the thermal resistance R' has the same dependence to γ as has the solution of Eq. (8). The knowledge of the resistance R'(γ=0), i.e. in the approximation of two interacting thermal baths, yields the exact resistance for the nanostructure configuration and for any values of γ according to the expression: R'(γ)=R'(γ =0)/(1-γ).

A simple estimation of R'(γ) is the resistance denoted R which is directly obtained from (R-$R_W$)/R=γ. This approximation is especially true for low or high Knudsen numbers. In the vicinity of Kn=1, a 11% disagreement was found in the case of the cylindrical contact. Finally, we can also infer that the correction (R-$R_W$)/R equals to the ratio between the contact cross section and the nanostructure surface at the ballistic limit.

**4. Conclusions**

In conclusion, we showed that the thermal resistance between a nanostructure and a half-body is augmented compared to the predictions of the half-body/half-body model. This deviation is mainly due to the multireflections of heat carriers inside the nanostructure. This increase depends on Knudsen number and on the ratio between the nanostructure and the contact surfaces. This contribution is significant when Kn>2. In the vicinity of Kn=1, we showed that the partially ballistic regime in the half-body also increases the contact resistance. The cases of the nanowire, the nanoparticle and the thin strip were calculated. The deviation to the current estimations reaches 500% at Kn=5 in the strip geometry. Temperature levels in metal tracks of integrate circuits might be strongly increased by this additional resistance. Highlighted effects also affect the thermal control of nanostructures, local probes and nanofabrication processes. We emphasize that the framework of our study is restricted to the particle phonon physics that implies a contact size larger than 10nm at ambient.

**APPENDIX I**

The calculation of the $\gamma$ coefficient is derived from the equations of the Matrix of Enclosure Theory presented in reference 16. This theory is basically derived from the heat flux balance on each surface. Considering an enclosure with N surfaces bounding a uniform isothermal medium at temperature $T_1$, it provides the net heat fluxes $q_j$ on surfaces j based on the following equation:

$$\sum_{j=1}^{N}\left(\frac{\delta_{kj}}{\varepsilon_j} - \frac{\rho_j}{\varepsilon_j}\tau_{kj}\right)q_j = \sum_{j=1}^{N}\left(\delta_{kj} - \tau_{kj}\right)q_j^b - a_{kj}q_g \qquad (I.1)$$

where $q_g$ is the flux emitted by the phonon gas and the geometric-mean transmittance $\tau_{kj} = (\pi S_k)^{-1} \int_{\Omega(S_k),S_j} e^{-\frac{r}{\Lambda}} \mathbf{u}.d\mathbf{S}_j d\Omega$ is the transmittance and $a_{kj} = (\pi S_k)^{-1} \int_{\Omega(S_k),S_j} \left(1 - e^{-\frac{r}{\Lambda}}\right) \mathbf{u}.d\mathbf{S}_j d\Omega$ is the absorbance. The surfaces are assumed to be diffuse and the emission in the medium is isotropic. $\rho$ is the reflection coefficient and $\varepsilon_j$ is the ratio between the phonon flux emitted by 420 the surface j and the phonon flux emitted if the surface were a perfect phonon emitter. The superscript b indicates the equilibrium (or blackbody) emission. We firstly calculate the temperature $T_1^l$ corresponding to the heat flux leaving the surface 1. Developing Eq. (I.1) when k=1 yields:

$$\left(\frac{1}{\varepsilon_1} - \frac{\rho_1}{\varepsilon_1}\tau_{11}\right)q_1 - \frac{\rho_D}{\varepsilon_D}\tau_{1D}q_D = -q_g(a_{11} + a_{1D}) + (1-\tau_{11})q_1^b - \tau_{1D}q_D^b \quad (I.2)$$

$\rho_D$=0 because the half-body absorbs all the phonons crossing the contact towards its direction. The surface 1 is assumed to be a non-emitting surface and $q_1^b$=0. The flux $q_1^l$ leaving the surface 1 is related to the net heat flux $q_1$ according to:[16]

430

$$q_1 = \frac{\varepsilon_1}{\rho_1}(q_1^b - q_1^l) = -\frac{\varepsilon_1}{\rho_1}q_1^l. \quad (I.3)$$

Combining Eqs. (I.2) and (I.3) leads to:

435 $(1-\tau_{11})q_1^l = q_g(a_{11} + a_{1D}) + \tau_{1D}q_D^b. \quad (I.4)$

$\rho_1$=1 because phonons are not absorbed on the nanostructure surface. The phonon energy is considered as fully reflected on the nano-object surface because the boundaries of the structure are free. Following Eq. (2), we consider that $q_1^l$, $q_D^b$ and $q_g$ are proportional to $T_1^l$, 
440 $T_0$ (the surface D is transmitting the phonons from the thermal bath) and $T_1$ respectively. Rewriting Eq. (I.4) yields:

$$T_1^l - T_0 = \frac{T_1[1-\tau_{11}-\tau_{1D}] + [-1+\tau_{1D}+\tau_{11}]T_0}{(1-\tau_{11})}, \quad (I.5)$$

because $a_{kj} = \alpha_{kj}(1-\tau_{kj})$ and $\alpha_{11} + \alpha_{1D} = 1$. Finally, it turns out that:

$$\frac{T_1^l - T_0}{T_1 - T_0} = \frac{[1 - \tau_{11} - \tau_{1D}]}{(1-\tau_{11})} = 1 - \gamma', \tag{I.6}$$

with $\gamma' = \frac{\tau_{1D}}{(1-\tau_{11})}$. To obtain the incident flux on surface D noted $q_D^i$, Eq. (I.2) is written with k=D:

$$\tau_{D1} q_1^l + \left(\frac{1}{\varepsilon_D} - \frac{\rho_D}{\varepsilon_D}\tau_{DD}\right) q_D = -q_g(a_{D1} + a_{DD}) - \tau_{D1} q_1^b + (1-\tau_{DD}) q_D^b. \tag{I.7}$$

Setting $\rho_D=0$ and $q_1^b=0$ again raises the following equation:

$$\tau_{D1} q_1^l + q_D = -q_g a_{D1} + q_D^b, \tag{I.8}$$

The simplification arises because D is a flat surface in such a way that $a_{DD}$ and $\tau_{DD}$ cancel. The definition of the net heat flux $q_D = q_D^l - q_D^i$ and the equality $q_D^b = q_D^l$ yield:

$$q_D^i = q_g a_{D1} + \tau_{D1} q_1^l = q_g + \tau_{D1}(q_1^l - q_g), \tag{I.9}$$

The configuration factor $\alpha_{D1}$ was also set to one because all the phonon flux emitted by the surface D inside the nanostructure reaches the surface 1. Replacing the fluxes by the corresponding temperatures leads to:

$$T_D^i - T_0 = (T_1 - T_0) + \tau_{D1}\left[(T_1 - T_0)(1-\gamma') + T_0 - T_1\right] \tag{I.10}$$

or

$$T_D^i - T_0 = (T_1 - T_0)(1 - \tau_{D1}\gamma'). \tag{I.11}$$

The final expression of $\gamma$ arises as: $\gamma = \tau_{D1}\gamma' = \frac{\tau_{D1}^2}{1-\tau_{11}}$.

The temperature $T_a$ is the average of the temperatures $T_0$ and $T_D^i$:

$$\frac{T_a}{T_1 - T_0} = \frac{T_D^i + T_0}{2(T_1 - T_0)} = \frac{1}{2}\left(1 - \gamma + 2\frac{T_0}{T_1 - T_0}\right), \tag{I.12}$$

and finally $\frac{T_a - T_0}{T_1 - T_0} = \frac{1-\gamma}{2}$.

**APPENDIX II**

465  When the nanostructure is a strip of infinite length, of width D and of thickness e=D/10 then $\tau_{11}$ equals to zero because the surface of the nanostructure is mostly flat, and γ reduces to $\tau_{1D}^2$. The configuration factor $\alpha_{1D}$ is equal to one in such a way that $\tau_{1D}$ can be written as (1-$L_{1D}/\Lambda$). Reference 16 directly provides the geometric mean beam length $L_{1D}$=0.175 D which leads to γ =(1-0.175/Kn)$^2$.

470  Following the same procedure, we solve the case of the horizontal wire of square section of edge D. The configuration factors are deduced from the reciprocity and the summation condition in the nanostructure: $\alpha_{D1}$=1, $\alpha_{1D}$=$S_D/S_1$=1/3, $\alpha_{11}$=1-$\alpha_{1D}$=2/3. The algebra of the mean beam lengths allows for writing:

$$\alpha_{D1}L_{1D} = \alpha_{Da}L_{aD} + \alpha_{Db}L_{bD} + \alpha_{Dc}L_{cD}, \tag{II.1}$$

475  where the indexes a, b, c refer to the three facets of the wire, the surface $S_b$ being parallel to the surface $S_D$. Due to the symmetry, $L_{aD}$=$L_{cD}$ and Eq. (II.1) reduces to $L_{1D} = 2\alpha_{Da}L_{aD} + \alpha_{Db}L_{bD}$. Decomposing the mean beam length $L_{11}$ leads to $S_1\alpha_{11}L_{11} = 2(2S_a\alpha_{ab}L_{ab} + S_a\alpha_{ac}L_{ac})$. We used the fact that $L_{ii}$=0 when i=a, b or c –because $S_a$, $S_b$ and $S_c$ are flat surfaces- and the reciprocity imposes that $L_{ij}$=$L_{ji}$. We finally end up with

480  $L_{11} = 2\alpha_{ab}L_{ab} + \alpha_{ac}L_{ac} = L_{1D}$. Using $L_{ab}$=0 and $L_{ac}$= δ D with δ =0.5588, the γ factor can be written as: $\gamma = \frac{(1-\delta/Kn)^2}{9 - 6(1-\delta/Kn)}$.

If the structure is a vertical wire of square section of edge D and length 10D, then $\alpha_{1D}=1/40$ and $\alpha_{11}=39/40$. The mean beam lengths between two opposite rectangles and between rectangles at right angles provide: $\gamma \approx \dfrac{\alpha_{1D}^2 (1-\delta/Kn)^2}{\delta'/Kn}$ where $\delta=3.467$ and $\delta'=1.059$.

485  For the cube of edge D, it is possible to show that $L_{11}=L_{1D}$ again in such a way that $\gamma = \dfrac{(1-\delta/Kn)^2}{25-20(1-\delta/Kn)}$ because $\alpha_{1D}=1/5$ and $\alpha_{11}=4/5$. The geometric-mean beam length coefficient is here $\delta= 0.6668$.

All the configuration factors were found in reference 16.

**CAPTIONS**

Table 1: The correcting coefficient γ is reported in Table 1 as a function of the Knudsen number, the geometric-mean beam length coefficient $\delta=L_{1i}/D$ and the structure shape. The Knudsen number is defined by the ratio between the phonon mean free path and the characteristic length D. The strip has a thickness e that is equal to the width D divided by 10. The wires have square sections of edge D. The cube has also an edge of length D.

Figure 1a: Schematic of the situation where the contact cross section is very small compared to the characteristic size of the nanostructure. The heat carriers are trapped inside the structure and they are thermalized. Grey stars represent a phonon-surface scattering event. The nanostructure can be assimilated to a perfect phonon absorber or a thermal bath. But the contact size has to be much smaller than 500nm in such a way that the particle phonon physics does not apply anymore.

Figure 1b: Schematic of the nanostructure/half body configuration. The characteristic dimensions of the contact cross section D, of the nanostructure L and of the phonon-phonon mean free path Λ are reported. D has to be larger than 10nm for the particle phonon physics to be applied. The nanostructure size L<500nm is hence on the same order of magnitude than D. The phonon mean free path in dielectric and semi-conductor crystals is also of the order of a few tens of nanometers. In this situation, multireflections occur and have to be taken into account.

Figure 2: Schematic of the different regimes in the nanostructure/half body case. L and D are the nanostructure and contact sizes. Λ is the phonon mean free path and $\lambda_{max}$ represents the wavelength of the predominant thermal phonons. The Wexler formula (dotted background) or 540   Eq. (1) is not adequate to describe the thermal contact between a nanostructure and a surface. Wexler formula requires that L>>D or L>>Λ. Our work treats the nanostructure case in the frame of the phonon-particle physics (no phonon diffraction) including multireflections in the nanostructure.

545   Figure 3a: Difference between the thermal resistance R of Eq. (7) and the Wexler resistance $R_W$ of Eq. (1) divided by the resistance $R_W$ as a function of the Knudsen number. The partially ballistic regime in the half-body is neglected. The cases of the strip structure, the wire of square section, the wire and the cube are reported.

550   Figure 3b: Evolution of the shape factor $\gamma=\delta R/R$ as a function of the Knudsen number for four different structures described in Fig. 3a.

Figure 4: The ratio $R/R_F$ x (1-γ) versus the Knudsen number when γ=0 and γ=1/2. The multireflections in the nanostructure and its shape are taken into account in the factor γ. The
555   black circles correspond to the mean of all calculated values for a given Knudsen number. The dashed line is a polynomial interpolation.

Table 1

|  | Strip | Wire ⊥ Half Body | Wire // Half Body | Cube |
|---|---|---|---|---|
| δ, δ' | 0.175 | 3.467, 1.059 | 0.5588 | 0.6668 |
| γ | $\gamma = (1-\delta/Kn)^2$ | $\gamma \approx \dfrac{\alpha_{1D}^2 (1-\delta/Kn)^2}{\delta/Kn}$ | $\gamma = \dfrac{(1-\delta/Kn)^2}{9 - 6(1-\delta/Kn)}$ | $\gamma = \dfrac{(1-\delta/Kn)^2}{25 - 20(1-\delta/Kn)}$ |